\documentclass{article}
\usepackage{cite}
\usepackage{amsmath,amssymb,amsfonts}
\usepackage{graphicx}
\usepackage{algorithm}  
\usepackage{algpseudocode}  
\usepackage{textcomp}
\usepackage{xcolor}
\usepackage{bm}
\usepackage{float}
\usepackage{hyperref}
\usepackage{flushend}
\usepackage{url}

\usepackage{mathrsfs}

\usepackage{comment}
\newtheorem{definition}{Definition}[section]
\usepackage{algorithmicx}  
\usepackage{xcolor}
\usepackage{amsmath} 
\usepackage{algorithm}  
\usepackage{algorithmicx}  
\usepackage{algpseudocode}  

\usepackage{arxiv}

\usepackage[utf8]{inputenc} 
\usepackage[T1]{fontenc}    
\usepackage{hyperref}       
\usepackage{url}            
\usepackage{booktabs}       
\usepackage{amsfonts}       
\usepackage{nicefrac}       
\usepackage{microtype}      
\usepackage{lipsum}		
\usepackage{graphicx}
\usepackage{natbib}
\usepackage{doi}

\title{WebFed: Cross-platform Federated Learning Framework Based on Web Browser with Local Differential Privacy}

\author{ {\hspace{1mm}Zhuotao Lian} \\
    School of Computer Science\\ and Engineering\\
	The University of Aizu,\\
Aizuwakamatsu, Japan \\
	\texttt{zhuotaolian@ieee.org} \\
	\And
{\hspace{1mm}Qinglin Yang}\thanks{Qinglin Yang is the corresponding author.} \\
College of Engineering \\
	Intelligent Transportation System,\\
	Sun Yat-sen University\\
	Guangzhou City, China \\
	\texttt{yangqlin6@mail.sysu.edu.cn} \\

	\And
{\hspace{1mm}Qingkui Zeng} \\
School of Electronics \\
	and Information Engineering,\\
	Nanjing University of Information\\
	Science \& Technology\\
	Nanjing, China \\

	\texttt{zenghuh1996@gmail.com} \\	

	\And
{\hspace{1mm}Chunhua Su}\\
    School of Computer Science\\ and Engineering\\
	The University of Aizu,\\
Aizuwakamatsu, Japan \\
	\texttt{chsu@u-aizu.ac.jp} \\

}

\hypersetup{
pdftitle={A template for the arxiv style},
pdfsubject={q-bio.NC, q-bio.QM},
pdfauthor={David S.~Hippocampus, Elias D.~Striatum},
pdfkeywords={First keyword, Second keyword, More},
}

\begin{document}
\maketitle

\begin{abstract}
For data isolated islands and privacy issues, federated learning has been extensively invoking much interest since it allows clients to collaborate on training a global model using their local data without sharing any with a third party. However, the existing federated learning frameworks always need sophisticated condition configurations (e.g., sophisticated driver configuration of standalone graphics card like NVIDIA, compile environment) that bring much inconvenience for large-scale development and deployment. To facilitate the deployment of federated learning and the implementation of related applications, we innovatively propose WebFed, a novel browser-based federated learning framework that takes advantage of the browser's features (e.g., Cross-platform, JavaScript Programming Features) and enhances the privacy protection via local differential privacy mechanism. Finally, We conduct experiments on heterogeneous devices to evaluate the performance of the proposed WebFed framework.
\end{abstract}


\section{Introduction}
During the past ten years, the breakthrough of machine learning (ML) technology has recently yielded encouraging results due to the development of computing power and big data, showcasing Artificial Intelligence (AI) systems able to assist life, education, and work in a variety of scenarios.

However, the conventional training mode encounters the issues, such as privacy leakage of user data and massive data transmission since it needs to gather all the raw training data. Hence, the privacy-preserving issue has attracted more and more attention \cite{house2012consumer}, and it has promoted a novel machine learning technology, which is called federated learning (FL). It leaves the personal training data distributed on their local devices and enables the clients to learn a shared global model by aggregating their local training results. It is seemed to be one instance of the more general approach to address the fundamental problems of privacy, ownership, and data isolated islands \cite{bonawitz2019towards}.

The emergence of FL has promoted the development of machine learning applications in many fields since it could solve the problem caused by the contradiction between model training and user data privacy, that is, collecting more user data is conducive to the training effect but may lead to user privacy leakage.

\begin{figure}[tbp]
    \centering
    \includegraphics[width=0.8\linewidth]{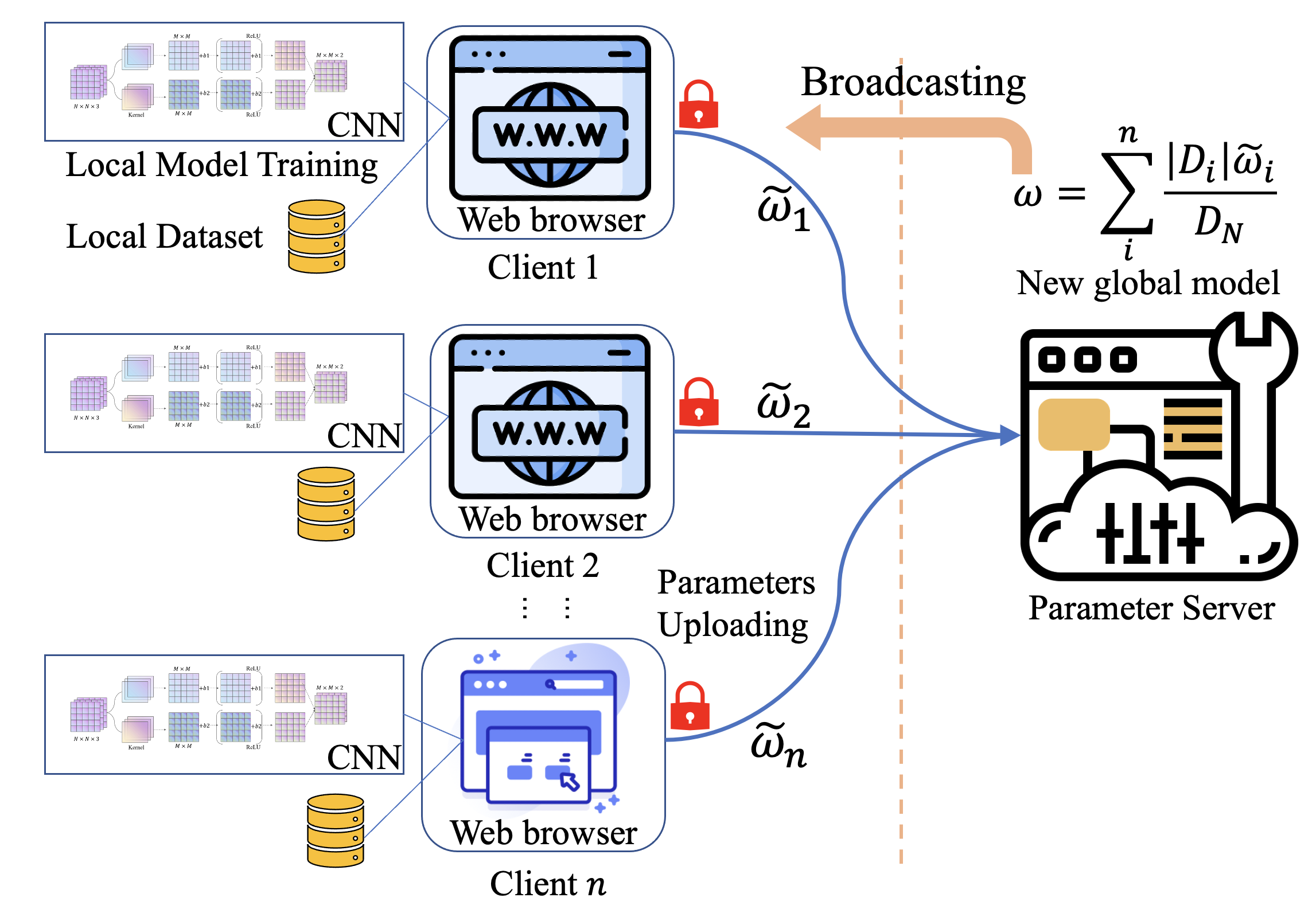}
    \caption{WebFed: Browser-based Federated Learning framework}
    \label{lbl:systemModel}
\end{figure}

Nevertheless, the most existing federated learning frameworks require the participants who have the heterogeneous operating systems (Windows, macOS, iOS, and Android) to install related software and configure complicated learning environment which seriously blocks their applications. Therefore, it is necessary to develop solutions other than standard methods for large-scale machine learning, distributed optimization, and other scenarios\cite{li2020federated}. That's is why we will try to develop a browser-based federated learning framework.

As an engine for developing distributed machine learning models and algorithms, the browser has the following advantages: 
(1) Browsers with cross-platform features and web programming languages make the software compatibility of machine learning models and algorithms obvious: nearly almost all computing devices can participate in machine learning training by contributing their computing resources to the entire learning process without any other software installation and could, with the same code, utilize the predictive models on the same devices. (2) WebGL and related technologies can make better use of integrated graphics cards to accelerate deep learning (DL) tasks, without the need for complex driver configuration of discrete graphics cards like NVIDIA that requires native DL framework \cite{ma2019moving}. (3) Browsers have great potential, are inexpensive and large in scale, and can widely bring complex ML learning and prediction to the public \cite{meeds2015mlitb}. 

Moreover, even FL has avoided the gathering of user's data, it still encounters the challenges on the privacy issues that the clients' information could be leaked by analyzing their uploaded parameters, such as the trained weights in deep neural networks (DNNs) \cite{9069945}. Comprehensive privacy analysis of deep learning models was firstly performed in \cite{nasr2019comprehensive}. They designed white-box inference attacks and tested the privacy leakage through the fully trained model's parameters and the updates during the whole training process. They also investigated the reasons for the information leakage of the deep learning models in the training data. To address the aforementioned information leakage issues, differential privacy (DP) is used as an element formulation of privacy in probabilistic terms to prevent privacy leakage of the information contained in its inputs by adding noise in centralized server \cite{fioretto2020differential}.



The above study motivates us to develop a browser-based cross-platform FL framework with local differential privacy that is capable of performing large-scale collaborative training on heterogeneous devices and enhances privacy protection by adding artificial noise. To our knowledge, it is the state-of-the-art exploration of developing the privacy-enhanced browser-based FL framework.

The contributions of this paper compose of three folds:
\begin{enumerate}
\item We firstly propose WebFed, a browser-based cross-platform federated learning framework in which machine learning models could be trained locally in web browsers with the clients' own local data. 
\item To strengthen privacy protection, we apply local differential privacy mechanism in WebFed. Before uploading the training results, each client will add artificial noise to local models' weights. Doing so could counter inference attacks without significantly negatively affecting performance.
\item The experiments on heterogeneous devices are conducted to evaluate the proposed WebFed framework. It is very easy to deploy because of its cross-platform nature and the results demonstrate that WebFed could achieve good training results even it runs in web browsers.
\end{enumerate}

The rest of this paper is organized as follows: Sec. \ref{lbl:relatedWork} summarizes the related work. Sec. \ref{lbl:framework} details the design of WebFed. Sec. \ref{lbl:experiment} evaluates the browser-based training performance and discusses the experiment results. In Sec. \ref{lbl:conclusion}, we give the conclusion of our current work. And the future direction is talked about in Sec. \ref{lbl:futureWork}.

\section{Preliminaries}\label{lbl:relatedWork}

\subsection{Machine Learning in The Browser}
TensorFlow.js is a library developed by Google to build and execute machine learning algorithms in JavaScript, which means the training process could be realized in the browser based on Node.js environment \cite{smilkov2019tensorflow}. It not only provides deep learning for JavaScript developers but also provides applications of deep learning in web browsers that support WebGL \cite{kletz2021open}. Recently, a browser-based ML application development tool was proposed to reduce the programming pressure of researchers \cite{ozan2021novel}. 

WebSocket provides full-duplex communication channels as a communication protocol on a single TCP connection. It is seemed to be able to reduce communication overhead and offer stateful, efficient communication among web servers and clients and thus could be a good approach for providing real-time information\cite{ogundeyi2019websocket}. 

With the breakthrough of hardware and communication technology, training on mobile devices has become possible. Due to recent advancements in the mobile web, it has enabled features previously only found in natively developed applications \cite{biorn2017progressive}. Hence, we strongly believe that more and more browser-based mobile machine learning applications will appear soon.


\subsection{Privacy Issues in Federated Learning }
Federated learning allows training a shared global model with a central server without gathering the clients' private data. By designing a privacy-preserving and bandwidth-efficient federated learning system, an application for the prediction of in-hospital mortality was developed in \cite{kerkouche2021privacy}. Moreover, FFD (Federated learning for Fraud Detection) was proposed to use behavior features distributed on the banks' local database to train a fraud detection model instead of the traditional fraud detection system that needs centralized data \cite{yang2019ffd}. In order to realize distributed machine learning with privacy protection, federated learning has shown great potential as it avoids the direct raw data sharing \cite{li2019asynchronous}.

However, it is still possible to leak privacy which is defined as the information an adversary can learn from the model about the users even it does not need to gather their private data \cite{fredrikson2015model, hitaj2017deep}. For example, an adversary with BlackBox or Whitebox access to the target model aims to determine whether a given target data point belongs to the private training data of the target model \cite{nasr2019comprehensive}. It's more effective for membership inference attacks against deep neural networks \cite{nasr2019comprehensive} since such models can better remind their training data with their immense capacities. 

\subsection{Local Differential Privacy}
Differential privacy is an element formulation of privacy in probabilistic terms. It can provide a strict preserve for an algorithm to prevent leakage of private information contained in its inputs for the centralized architecture \cite{fioretto2020differential}.
Local differential privacy (LDP) allows the clients to perturb their information locally to provide plausible deniability of their information without the necessity of a trusted party \cite{8731512}. In \cite{8731512}, the authors exploited novel LDP mechanisms to collect a numeric attribute. In terms of worst-case noise variance, it usually has a better performance on accuracy than existing solutions. In \cite{kairouz2014extremal}, they focus on the fundamental trade-off between local differential privacy and information-theoretic utility functions, because data providers and analysts want to maximize the utility of statistical inferences performed on the released data. In \cite{8640266}, the authors propose to apply LDP to protect the clients' data by distribution estimation. In our WebFed, we enhance privacy protection by applying local differential privacy to add artificial noise to the training results before uploading to the parameter server.

\subsection{Towards WebFed}

Significantly different from existing works, we are the first to develop a browser-based federated learning framework. Based on the restrictions on the development and deployment of federated learning mentioned above, as well as the emergence of browser-based machine learning libraries, and the cross-platform characteristics of browsers, we propose WebFed, a browser-based federated learning framework with local differential privacy to facilitate future research related to federated learning, as well as the deployment and implementation of real-world applications.

\section{Framework Design}\label{lbl:framework}
In this section, we first describe the architecture design of WebFed, explain the Workers and their relationships, and then we will introduce the details of the WebFedAVG algorithm.

\begin{figure}[tbp]
    \centering
    \includegraphics[width=0.5\linewidth]{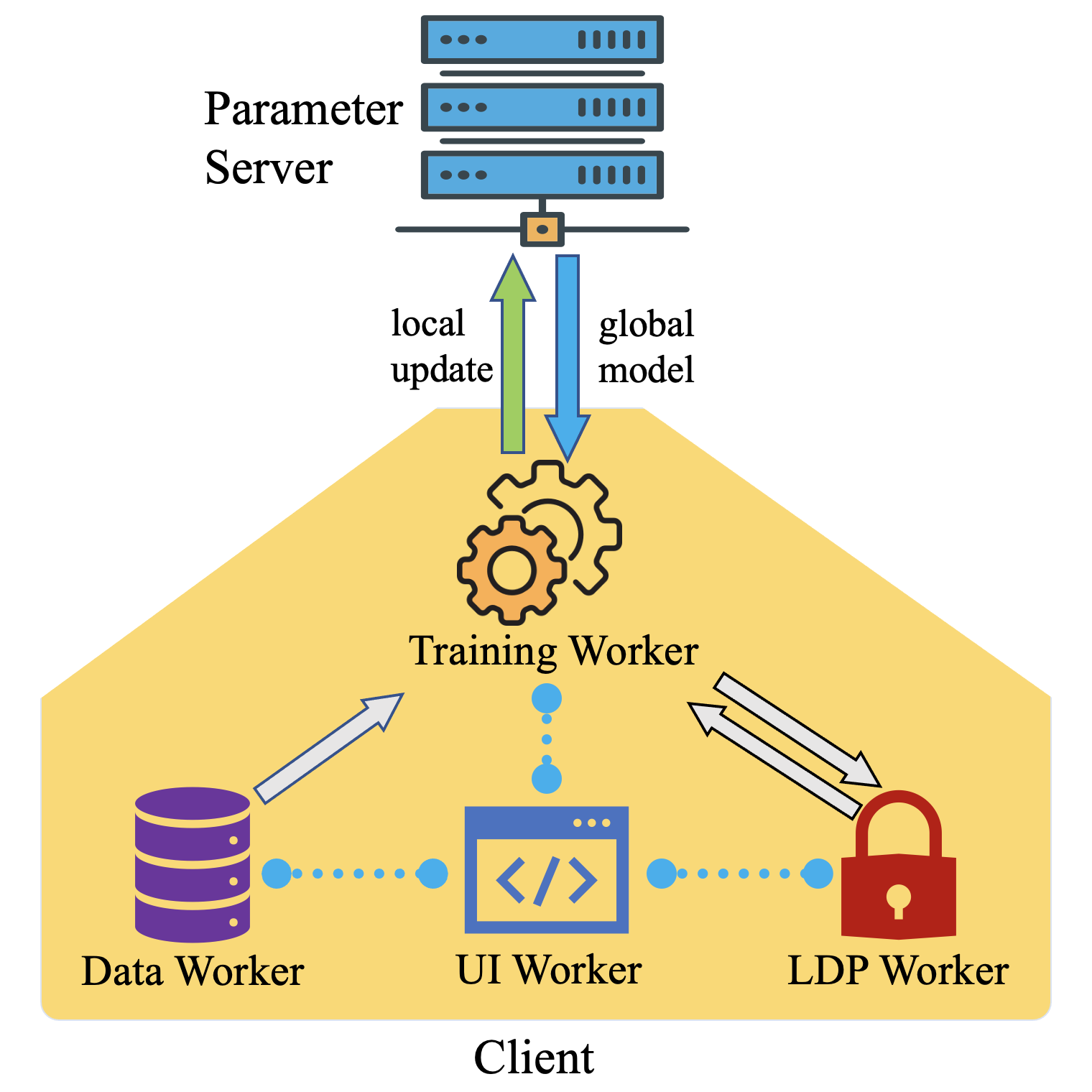}
    \caption{Schematic Illustration of WebFed architecture. Icons in this figure are from \textit{https://www.flaticon.com/}.
   } \label{lbl:tech}
\end{figure}

\subsection{Overall Design of WebFed}
Under WebFed, the Browser-based federated learning framework, a shared deep neural network (DNN) model is jointly trained in browsers among $N$ clients without gathering their private data.  We assume there is a parameter server that is in charge of aggregating the clients' updates (weights of their local models)  by convolutional methods (e.g., FedAvg, FedSGD) and broadcasting the new global model. Parameter Servers (PS) are \textit{Node.js} applications that control the communication between devices and hosts ML projects. 

As depicted in Fig. \ref{lbl:tech}, the clients in WebFed could be viewed as the composing of four Workers, that is:

\begin{itemize}
    \item \textbf{UI Worker:} 
    When the corresponding web page is opened, UI Worker is initialized. Clients can perform some intuitive operations through UI Worker, such as connecting to the server, selecting training tasks, and so on. UI Worker is mainly responsible for the interaction with the user and the specific training configuration on the browser side.
    \item \textbf{Data Worker:}
    Data Worker participates in training mainly by managing local datasets. It could optimize the entire training process by pre-loading and pre-processing the training data.
    \item \textbf{LDP Worker:}
    LDP Worker is in charge of adding artificial noise to the local training results before global aggregation. It accepts the model from the Training Worker and returns the perturbed model weights.
    \item \textbf{Training Worker:}
    Training Worker is responsible for performing specific training tasks and interacting with the server. First, it will receive the global model broadcasted by the server and then update the maintained local model. It accepts the training data provided by Data Worker and then conducts the in-browser training process. After the local training, it will then interact with the LDP Worker as Fig. \ref{lbl:tech} illustrates to obtain the model weights with artificial noise and upload them to the parameter server for global aggregation.
    \end{itemize}




\subsection{Algorithm Design}
\subsubsection{Federated learning tasks}
Let  $D_i$ denote the local training dataset of client $i$, where $i \in  N$ and  we consider the number of all training data samples is denoted as $D_{N}$, and $\bigcup_{i} D_{i}=D_{N}$. For the $j$-th training sample of client $i$ in a classification problem, it includes input training data and corresponding labels. The full parameters of the training model are denoted as $\bm{\omega}$ and the loss of $j$-th data sample is denoted as $l_j( \bm{\omega};(x_{j},y_{j}))$, shorted as $l_j(\bm{\omega})$. The training goal is to minimize the loss $L(\bm{\omega} )$ that is derived from Eq. (\ref{fwi}) and Eq. (\ref{fw}). Assuming that each client has its own training data $D_i$, then the local loss is expressed as:
\begin{equation}\label{fwi}
L_i(\bm{\omega})=\frac{\sum_{j=1}^{|D_i|}l_j(\bm{\omega}) }{|D_i|},\ {\rm for}\ j \in |D_i|.
\end{equation}
Therefore, the global loss is then calculated as:
\begin{equation}\label{fw}
L(\bm{\omega})=\frac{\sum_{i=1}^{N}{|D_i|}L_i(\bm{\omega}) }{D_{N}}.
\end{equation}
where $n$ to denote the total number of clients participating in the training process. For updating model parameters under such a distributed architecture, we utilize the classic gradient-descent methods (e.g., Adam, ASGD) which is mathematically expressed below:
\begin{equation}
 \bm{\omega}_i(t)=\bm{\omega}_i(t-1)-\eta\nabla L_i(\bm{\omega}_i(t-1)).
\end{equation}
where $t$ denotes the index of training epoch, and $\eta$ denotes a step size in the gradient descent method. 

\begin{algorithm}[t]
\small
\caption{WebFedAVG} 
\label{alg1}
\begin{algorithmic}[1]
\State{\textbf{Input}: $\bm{\omega}_0$, $\epsilon$, m, T, $D_{i}$, and $\eta$ 
 }
 \State{\textbf{Output}: $\bm{\omega}$}
\State Register all available clients on the server side
\State Initialize all clients with model $\bm{\omega}_0$ 
\For{round $t=1, 2, ..., T$}
    \State Server randomly selects $m$ clients
    \For{each client $i=1,2,...,m$}
        \State \textbf{In-browser Training:}
        \State $\bm{\omega}_i(t) \leftarrow \bm{\omega}_i(t-1)-\eta\nabla L_i(\bm{\omega}_i(t-1))$
        
        \State \textbf{LDP Process:}
        
        \State $\tilde{\bm{\omega}}_{i}(t)=\bm{\omega}_{i}(t) + Lap^{i}(\Delta s/\epsilon)$
        \State Upload $\tilde{\bm{\omega}}_{i}(t)$ to server
    \EndFor

    \State{The server receives all updates and does}
        \State \textbf{Global Aggregation:}
        \State $\bm{\omega}(t)\leftarrow 
        \sum_{i}  \frac{|D_{i}| {\tilde{\bm{\omega}}}_{i}(t)}{\sum_{i} |D_{i}|}$

    \State{The server sends new global model $\bm{\omega}(t) $ to all clients}    
    \For{each client $i=1,2,...,n$}
    \State{$\bm{\omega}_i(t) \leftarrow \bm{\omega}(t)$. }
    \EndFor
\EndFor
\end{algorithmic}
\end{algorithm}

\subsubsection{Local Differential Privacy}
In this subsection, we firstly introduce the related definition on $\epsilon-$LDP and then we demonstrate how to apply it into our framework.

WebFed guarantees a certain degree of privacy as the clients never send their private raw data to the parameter server publicly. However, a little parameter value could still be inferred from the shared updates in the local model, thereby further leveraging privacy-preserving mechanisms is necessary.

$\epsilon-$LDP inherits the features of centralized DP \cite{9069945} and is able to resist privacy attacks from the untrusted server, provides a strong criterion for privacy information preservation of distributed clients. Now, we formally define LDP as follows.
\begin{definition}[Local Differential Privacy]
An randomized mechanism $\mathcal{M}$ satisfies $\epsilon -$local differential privacy ($\epsilon-$LDP), where $\epsilon \geq 0$, if and only if for any input $v$ and $v^{\prime}$, and any output $y$ of $M$, we have 
\begin{equation}
\forall y \in \operatorname{Range}(\mathcal{M}): \operatorname{Pr}[\mathcal{M}(v) \in y] \leq e^{\epsilon} \operatorname{Pr}\left[\mathcal{M}\left(v^{\prime}\right) \in y \right]
\label{lbl:ldpTheroty}
\end{equation}
\end{definition}
where, $\operatorname{Range}(\mathcal{M})$ denotes the set of all possible outputs of the algorithms $\mathcal{M}$. The notation $Pr[\cdot]$ means probability. $\epsilon$ denotes the distinguishable bound of the probability distributions of neighboring inputs to achieve almost identical output results for any neighboring inputs, which is also called the privacy budget. 

Basically, LDP provides a stronger level of protection compared to the centralized DP setting, because each user only sends the perturbed report. Even in the case of a malicious parameter aggregator, the privacy of each client is still protected. According to the above definition, no matter what extent of background knowledge the aggregator holds, it cannot distinguish with high confidence whether the real tuple is $v$ or $v^{\prime}$ after receiving the perturbed report under LDP protection. Therefore, deploying LDP in our WebFed framework could further ensure that model weights are protected from attacks by an aggregator with background knowledge.

\subsubsection{WebFedAVG overall workflow}

Based on the above details, we briefly summarize the overall workflow of WebFedAVG. As shown in Algorithm \ref{alg1}, the parameter server will initialize the training model and broadcast it to all clients before the collaboratively training process, and it will randomly select several clients as participants in each round of training. 

When clients receive the initial global model, they will start their in-browser training with their own local data. For privacy concerns, they will add artificial noise to their training results as indicated in line 11 of Algorithm \ref{alg1}, and then upload them to the parameter server. 

The server will then do weighted averaging to generate a new global model and broadcast it to all clients. At the same time, the clients selected for the next round will receive a request along with the new global model.

The above procedure repeats until reaching the number of epochs or the accuracy requirement.

\section{Experiment and Anslysis}\label{lbl:experiment}
In this section, we mainly conduct experiments to evaluate the performance of the WebFed framework and analyze the experiment results.
\addtolength{\topmargin}{0.01in}
\subsection{Experiment Setup}

Here we choose a lightweight deep learning network, LeNet with two convolutional layers (kernel size 5$\times$5) and two max pooling layers (kernel size 2$\times$2), as the collaborative training model. Our evaluation is conducted on the MNIST that is the most popular benchmark dataset of Machine Learning models.



\begin{table}[tbp] 
\caption{training devices}
\label{lbl:mobileDevice}
\begin{center}
\begin{tabular}{|c|c|c|c|c|}

\hline
\textbf{Type} & \textbf{Name}& \textbf{Operating system}& \textbf{hardware}&\textbf{browser}\\
\hline
smartphone& iphone 12& ios15 & A14 Bionic& Chrome  \\
smartphone& HUAWEI & Harmony&KIRIN 980& Chrome  \\
PC& Dell &Ubuntu 18.04& i7 CPU & FireFox \\
PC& Dell &Ubuntu 18.04& i7 CPU & FireFox \\
PC &MAC &BigSur& M1& Safari \\
\hline
\end{tabular}
\label{tab1}
\end{center}
\end{table}

We deploy the parameter server on a PC. For clients, we select two smartphones and three PCs to test the WebFed framework, and the specific information of the devices are shown in table \ref{lbl:mobileDevice}. To simplify the experiment, we stipulate that all devices participate in the whole training process.

\subsection{Results and Analysis}

\subsubsection{Comparison between WebFed and FL}

\begin{figure}[t]
    \centering
    \includegraphics[width=0.7\linewidth]{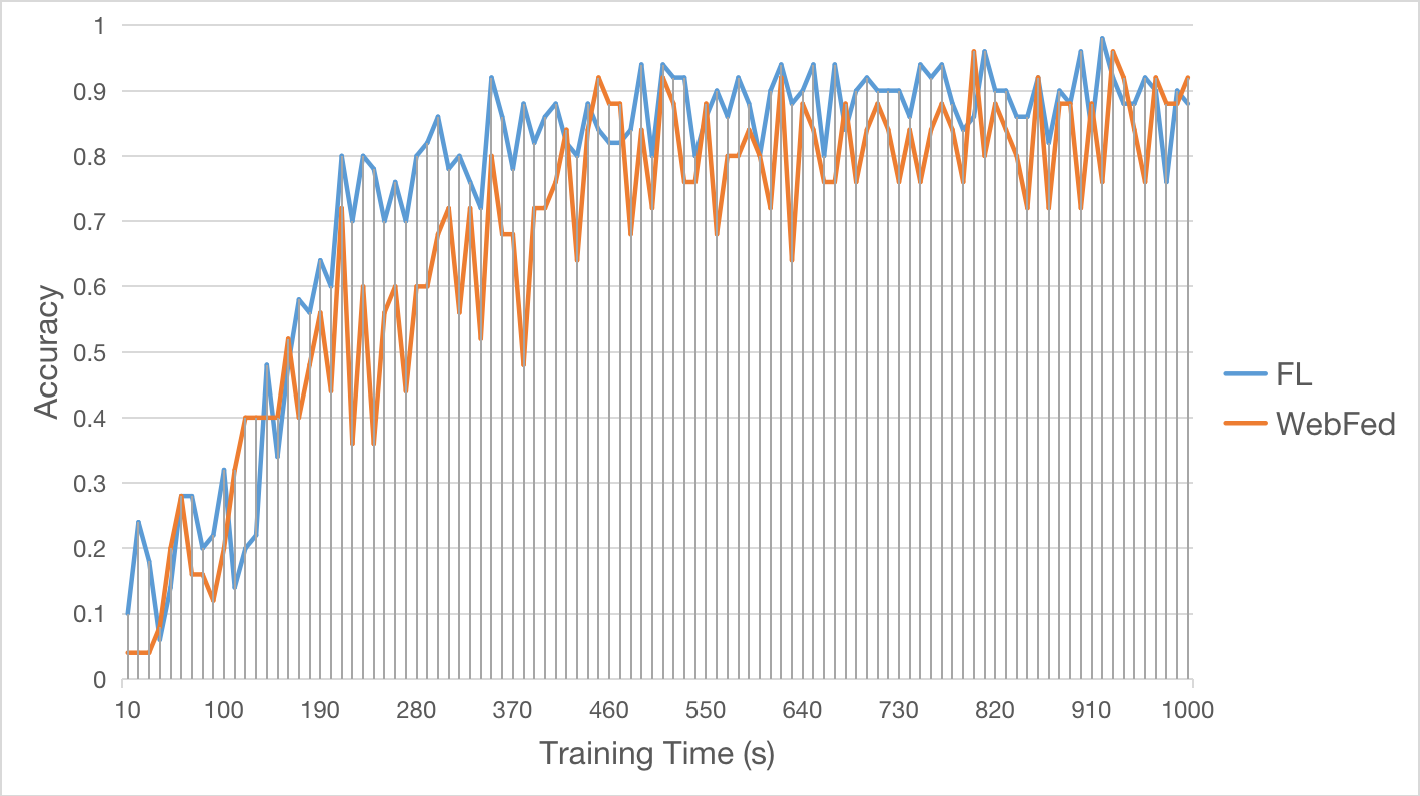}
    \caption{The accuracy of WebFed versus FL on MNIST.}
    \label{lbl:WebFedAccuracy}
\end{figure}

With the experiment settings, we obtain the learning trend for WebFed framework of both the accuracy and loss values. Moreover, we also test on convolutional federated learning framework that does not run on the browsers, which is called FL for short in the following. In order to control the experimental variables, we apply local differential privacy to both, and the privacy budget is 3. The experiment results are demonstrated in Fig. \ref{lbl:WebFedAccuracy} and Fig. \ref{lbl:WebFedloss}, both frameworks could achieve the convergence of training after fixed epochs. 

In Fig. \ref{lbl:WebFedAccuracy}, we compare the accuracy between FL and WebFed. And it illustrates that after the same training time, the accuracy of the traditional federated learning framework will be slightly higher than WebFed. We also can find that the curve trend of WebFed is always showing fluctuated than FL, when the training time is about 300 seconds, the difference in accuracy between FL and WebFed is obvious, the gap is about 13\%. But as the training continues, the gap between the two frameworks is getting smaller and smaller. 

 We think the reason for this context might be the result of the interaction of multiple factors (e.g., the approach for the selection of clients, the heterogeneity of clients, the quality of the dataset). Indeed, the performance of WedFeb is close to FL during training while the gap still exists.

\subsubsection{Comparison under different privacy budgets}
\begin{figure}[ht]
    \centering
    \includegraphics[width=0.7\linewidth]{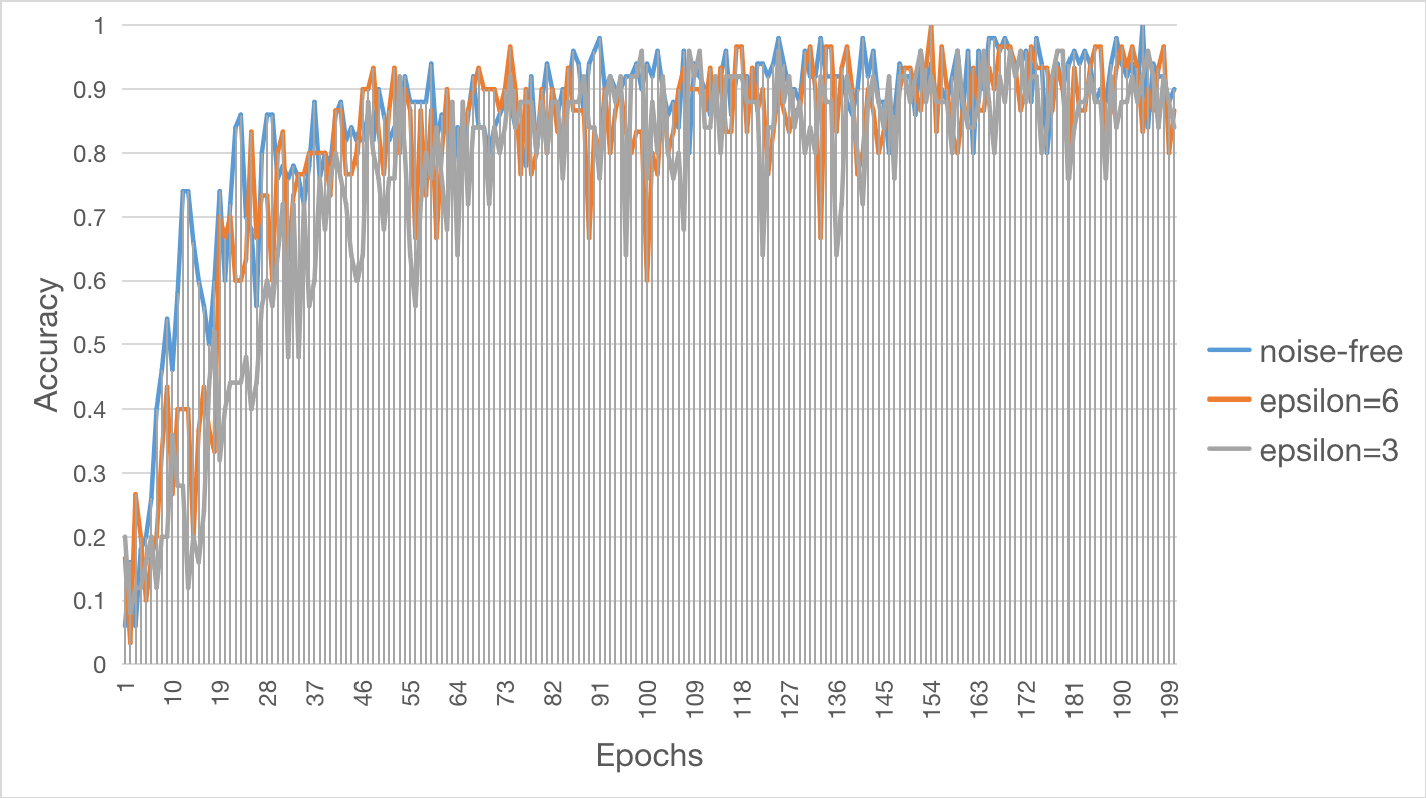}
    \caption{The accuracy of WebFed versus epsilon on MNIST.}
    \label{lbl:noiseAccuracy}
\end{figure}

To explore the impact of adding artificial noise on the training process, experiments are conducted by setting different privacy budgets. We consider three cases, namely, epsilon is equal to 3, 6, and noise-free (no noise is added). We train the shared model for 200 epochs, and Fig \ref{lbl:noiseAccuracy} shows the impact of different privacy budgets on global model accuracy. We can find that the accuracy of the model keeps getting lower with the privacy budget epsilon decreases. 

In the early and middle stages of the training, the gap among the three is the biggest. After 50 rounds of training, when epsilon is equal to 6 and noise-free, the accuracy is about 0.8. In the case of epsilon equals 3, the accuracy is the lowest, around 0.72. Compared with the other two cases, the accuracy gap is about 10\%. After 200 epochs of training, when epsilon is equal to 3 and 6, the accuracy is about 87\% and 90\%, respectively. It is obvious that it has the highest accuracy for noise-free, about 92\%. It is because the smaller the privacy budget, the higher the requirements for privacy protection, and the more affected the training effect. 

Through our experiments on different privacy budgets, setting an appropriate privacy budget can reduce the impact on the training performance while strengthening privacy protection at the same time.



\section{Conclusion}\label{lbl:conclusion}


In this paper, considering the difficulty of the cross-device deployment of federated learning, we proposed WebFed, a browser-based federated learning framework with privacy preservation. By leveraging the advantage of the cross-device characteristics of web browsers, WebFed is more conducive to actual federated learning research and application landing. What's more, in order to strengthen privacy protection, we apply local differential privacy mechanism to add artificial noise to the local updates before global aggregation. 

Finally, we conduct the experiments on heterogeneous devices (smartphones and PC) to evaluate the performance of WebFed. The experimental results show that our browser-based framework can achieve a training effect close to the conventional framework in terms of model accuracy. We also found that an appropriate privacy budget setting can enhance the privacy protection with only a subtle impact on the training performance.

\section{Future direction}\label{lbl:futureWork}






In the future, to further simplify deployment and increase the flexibility of federated learning, we plan to integrate the server into the browser. In this way, participants and servers (which could also be regarded as a participant) only need to manually select their roles in the browser and form a group for federated learning, rather than deploying servers separately. Furthermore, we will optimize corresponding mechanisms for clients' joining and leaving, and conduct more thorough experiments to evaluate the performance of WebFed. Machine learning in the browser is still in its infancy. We will further study the limitations of in-browser training, improve the existing work and then publish WebFed as an open-source framework.


\bibliographystyle{unsrtnat}
\bibliography{references}

\end{document}